# Observation of Magnetic Droplets in Magnetic Tunnel Junctions

Kewen Shi[1,†], Wenlong Cai[1, 2,†], Sheng Jiang[3, 4,†], Daoqian Zhu[1], Kaihua Cao[1, 2], Zongxia Guo[1], Jiaqi Wei[1], Ao Du[1], Zhi Li[1, 2], Yan Huang[1], Jialiang Yin[1, 2], Johan Åkerman[3, 4*], Weisheng Zhao[1, 2*]

[1]Fert Beijing Institute, School of Integrated Circuit Science and Engineering, Beijing Advanced Innovation Center for Big Data and Brain Computing, Beihang University, Beijing, China
[2]Beihang-Goertek Joint Microelectronics Institute, Qingdao Research Institute, Beihang University, Qingdao 266000, China
[3]Department of Physics, University of Gothenburg, 412 96, Gothenburg, Sweden
[4]Department of Applied Physics, School of Engineering Sciences, KTH Royal Institute of Technology, Electrum 229, SE-16440 Kista, Sweden

## ABSTRACT

Magnetic droplets, a class of highly non-linear magnetodynamical solitons, can be nucleated and stabilized in nanocontact spin-torque nano-oscillators where they greatly increase the microwave output power. Here, we experimentally demonstrate magnetic droplets in magnetic tunnel junctions (MTJs). The droplet nucleation is accompanied by a power increase of over 300 times compared to its ferromagnetic resonance modes. The nucleation and stabilization of droplets are ascribed to the double-CoFeB free layer structure in the all-perpendicular MTJ which provides a low Zhang-Li torque and a high pinning field. Our results enable better electrical sensitivity in the fundamental studies of droplets and show that the droplets can be utilized in MTJ-based applications.

MTJ-based Spin torque nano-oscillators (STNOs) have recently attracted wide interest in nanomagnetism and spintronics[1-4]. As broadband microwave signal generators, STNOs are able to generate signals with frequencies ranging from a few MHz up to hundred GHz[5-8], leading to potential applications in radio frequency electronics and artificial intelligence[9-12]. Nevertheless, achieving high output power is still a large challenge for conventional STNOs[7]. Magnetic and magnetodynamic structures[13-15], such as vortices[16], bullets[17, 18], and droplets[19], may provide one way to increase their power emission as they maximize the use of the available magnetoresistance thanks to their high precession amplitude. Especially, magnetic droplets are observed in nano-contact (NC) STNOs with a strong perpendicular magnetic anisotropy (PMA)[20, 21]. The microwave output power of the magnetic droplet mode was reported as 40 times higher than that of the normal ferromagnetic resonance (FMR) mode, mainly due to its large precession angle[7, 19, 20]. However, all experimental work on magnetic droplets has until now focused on spin-valve (SV) structures[19, 20, 22-24], and spin-hall nano oscillators (SHNOs)[25, 26]. The very low magnetoresistance (MR) (~1%) in SVs and SHNOs limits the power emission and hence any further use in STNO-based applications. Comparatively, magnetic tunneling junctions with strong PMA (pMTJs) have presented a high tunneling magnetoresistance (TMR)[27], reaching 249% especially for double-CoFeB free layer (DFL) pMTJs[28] which becomes the main structure in MTJs based MRAM. Therefore, one might expect to observe magnetic droplets in pMTJ based NC-STNOs. Nevertheless, our previous experiment shows that it is difficult to format a stable droplet in a single-free layer (SFL) MTJ[29], which may result from the large Zhang-Li torque induced by the lateral current. In contrast, the DFL pMTJs[30] are expected to suppress this large Zhang-Li torque and therefore benefit to format a stable magnetic droplet.

Here we experimentally observe and investigate stable magnetic droplets in a DFL pMTJ and find that the droplet enhances the emission power by more than two orders of magnitude with respect to the FMR mode precession in the same device. Furthermore, using micromagnetic simulations, we argue that the stable magnetic droplets in MTJs are mainly due to the combination of the low Zhang-Li torque and the strong pinning field in the DFL. Our findings provide a comprehensive understanding of the nucleation of magnetic droplets in MTJs. It will pave the way for further optimization for use of magnetic droplets in MTJs.

Figure 1a shows a schematic and stack information of our NC-STNO device. The film stack is composed of a [Co/Pt]-based pinned layer (PL), a CoFeB reference layer (RL) and a CoFeB/W/CoFeB coupled DFL, both with strong PMA. Figure 1b shows the out-of-plane (OOP) and in-plane (IP) magnetization hysteresis loops of a corresponding unpatterned film. The IP hysteresis loop shows a typical hard-axis response while the OOP hysteresis loop exhibits three distinct switching fields corresponding to the switching of the RL, DFL, and PL, respectively. The saturation magnetization $M_s$ of the DFL is about 987 kA/m. The PMA field $\mu_0 H_k$ is 120.9(8) mT (see Figure S1 in Supporting Information). A TMR ratio of 12.6% is measured in the fully processed STNO (Figure 1c).

After having confirmed the static behavior, we study their magnetodynamics in Figure 2. At a moderate external magnetic field sweeping from 100 to 25 mT with the direction $\theta_{ext} = 30°$, a significant high-frequency signal is observed as shown in Figure 2a, corresponding to the ordinary FMR-like mode. When the field is swept back to lower than 20 mT, it exhibits an abrupt frequency drop (~2.5 GHz) (see the details from the PSD spectra in the inset of Figure 2b) and a dramatic increase of the microwave power from 2 pW to nearly 600 pW. Simultaneously, the resistance jumps to an intermediate state between the values of the antiparallel (AP) and parallel (P) states (pink line in Figure 2a). An ~1 GHz dynamic signal with an enhanced power emission accompanied by the intermediate resistance state can be also observed in the angular dependent measurements in Figure 2c-d. Another significant feature of the droplet soliton is the low-$f$ noise, shown in Figure 2a, which is caused by its drift instability and subsequent re-nucleation[31]. We found that the low-$f$ noise, weak dependence of field and current, presents an obvious peak at ~200 MHz, which is far away from the 1/$f$ noise caused by the spin-transfer torque (STT) induced incoherent precession[32, 33]. Meanwhile, the ~GHz signal, low-$f$ noise could exist till -50 mT for -2.4 mA and 40 mT for 2.4 mA (see Figure S3 in supporting information) with an angle of 60 degrees, in which the magnetization of the free layer is supposed to be saturated. Figure 2e shows the perpendicular field-swept resistance measurements at different negative currents. As the dashed lines show, the saturated magnetic fields for DFL should be larger than 25 mT or less than -20 mT. However, it is obvious that the STT torques could form the intermediate resistance states (vanish over -60 mT) under saturated magnetic fields which are always accompanied by the low-$f$ noise shown in Figure 2g. Due to the collinear between the FL and RL magnetizations under perpendicular fields, no high-frequency signals is expected[31]. Moreover, the intermediate resistance states nucleation boundary shows a linear dependence on both current and field in Figure 2e, which are similar to the droplet boundary in SVs[34]. Since the saturated magnetic fields could suppress the domains rather than the droplets, all these observations lead to the conclusion that the observed new dynamic structures fully consistent with the nucleation of a magnetic droplet soliton.

Upon further characterization, we found another state which has an intermediate resistance but no low-$f$ noise. To further clarify the nature of this apparently non-dynamical state, we measured frequency spectra under perpendicular field sweeps at $I$ = -1.6 mA, and current sweeps at field of $\mu_0 H$ = -15 mT with an angle of $\theta_{ext}$ = 30°; shown in Figs. 3a-b, respectively. When the field is swept in the negative direction, the magnetization of the DFL stays AP to the RL's magnetization from 30 till -5 mT without STTST as shown in Figure 1c. However, since the STT from the negative applied current favors the AP state, the magnetization instead tends to maintain the original AP state underneath the NC area, while only the DFL area outside of the NC switches to the P state. This partial switching of the magnetization induces an intermediate resistance state, which is slightly higher than the resistance of the droplet and is completely void of any accompanying low-$f$ noise. This novel partially reversed state is hence not a precessing droplet, but more likely a "static magnetic bubble" with a reversed core, which is similar with SV-based NC-STNOs[35], where the current density is not enough to keep the droplet in a state of precession. The bubble state remains stable until the external magnetic field is swept down to -14 mT. By increasing the fields, the pressure on the static bubble makes it shrink until it again transforms into a precessing droplet with a step-like decrease of resistance and the reappearance of low-$f$ noise. The transformation from a static bubble to a precessing droplet may result from the STT becoming strong enough to again compensate for the damping torque as the magnetic field pushes the bubble perimeter into the NC region. Besides, the field sweep from negative to positive shows a similar property.

As for the current sweep in Figure 3b, the P state switches into a droplet state with a low-$f$ noise and step-like increase of the resistanceat a threshold current of $I_{th}$ = -1.57 mA. The small near ~1 GHz signals of the magnetic droplet also could be observed during the current sweep as shown in Figure 3b. The magnetic droplet is stable while the current is further decreased to -2.2 mA. When the current is swept back, the droplet still exists even though the current is much lower than the threshold current $I_{th}$ of nucleation of a droplet. As the low-$f$ noise damps out at around -0.9 mA, the dynamic droplet starts to slow down, and turns to a static bubble at further lower current with the disappearance of the ~1 GHz signal. The obvious hysteresis indicates the different magnitude of stimulus that is required for the droplet nucleation and annihilation[20].

We note that the threshold current density of droplet nucleation in our devices can be as low as ~5 MA cm$^{-2}$ with a magnetic field of ~20 mT which is quite less than that in the SVs (~130 MA cm$^{-2}$ under 0.25 T)[22]. Such low current density could be associated with *i*) slightly lower damping (0.02 for CoFeB here, 0.03 for [Co/Ni] multilayers[22]); *ii*) different spin-torque efficiency and *iii*) different device structures inducing different current distributions, and so on. We would also like to emphasize here that the lower threshold current density for droplets is critical in the MTJs with a power emission of up to 600 pW, providing a ~300 times enhancement (compared with the FMR-mode signal), which is more advantageous than that in SVs or SHNOs (see the chart in supplementary note 4).

In contrast to our previous work on an SFL MTJ[29], the magnetic droplets are successfully observed in a DFL MTJ here. To clarify the essential difference, we perform the micromagnetic simulations[23, 24] to further analyze the stability of magnetic droplets. Usually, the size of the droplet is determined by a combination of factors, such as the external magnetic field, the Zhang-Li torque induced by realistic lateral current spreading[22], and the pinning field of the free layer. For comparison, we calculated the current distribution in free layers of a SFL MTJ and an SV by COMSOL. The lateral current spread -$J_x$ is much higher in MTJs than that in SVs[22] due to the existence of the MgO barrier. Such high -$J_x$ induces a larger Zhang-Li torque in the free layer. Consequently, it further hinders the observation of stable droplets in MTJs. To explain the reason for stable droplets in our DFL MTJ, we have carried out the same current distribution calculation and a micromagnetic simulation as shown in Figure 4. In contrast, the -$J_x$ in DFL (usually with a MgO capping layer (CL)[28]) is near half of that in the SFL MTJs. It is associated with the current shunting at the MgO in the CL before entering the free layer, although this may consume a bit more current in the CL. For such a reason, it should be easier to obtain and stabilize the magnetic droplets in a DFL than an SFL. as seen in Figs. 4e and 4f, the region for a stable magnetic droplet is apparently wider in the DFL (-$J$ = 4.18 - 8.31 MA cm$^{-2}$) than that in a SFL (-$J$ = 3.5 - 4.3 MA cm$^{-2}$). The evolution of droplet nucleation or annihilation in SFL and DFL MTJs are displayed in Figs. 4(c) and 4(d).

Moreover, an inhomogeneous thin W insert layer between double-free layers can cause local DMI and non-uniform RKKY distribution, which induces a much higher pinning field compared with an SFL MTJs[30]. To figure out the influence of this pinning field, we performed a simulation for a DFL MTJ with RKKY distribution and DMI in Figure 4g. The range for nucleation of magnetic droplets becomes wider, indicating that the higher pinning field in our DFL MTJ highly stabilizes the observed magnetic droplets.

By using the DFL structure in pMTJs, a magnetic droplet is generated for significant power enhancement. In DFL structures, the low Zhang-Li torque and high pinning field have been identified to be responsible to stabilize the magnetic droplet. The high TMR of MTJs, compared to the low MR of SVs, enables one order of magnitude smaller current density to induce a comparatively higher power emission. Furthermore, the generation of magnetic droplets has been observed to increase the microwave power by at least two orders of magnitude compared to FMR-like modes. The droplet implementation in MTJs provides an effective method to improve the power of STNOs for radio frequency electronics. This new finding could launch an alternative path towards exploring applications of spintronic devices.

**Methods.**

**Sample fabrication.** The p-MTJ were deposited on a thermally oxidized Si substrate (300 nm $SiO_2$) by a Singulus TIMARIS 200 mm magnetron sputtering machine. The full stacks are composed of, from the substrate side, Ta(3)/Ru(20)/Ta(0.7)/Pt(1.5)/[Co(0.5)/Pt(0.35)]$_6$/Co(0.6)/Ru(0.8)/Co(0.6)/[Pt(0.35)/Co(0.5)]$_3$/Pt(0.25)/Ta(0.2)/Co(1.2)/W(0.25)/CoFeB(0.9)/MgO(~0.8)/CoFeB(1.2)/W(0.3)/CoFeB(0.5)/MgO(~0.8)/Pt(1.5)/Ta(3)/Ru(7) (thickness in nanometers). The p-MTJ film was annealed at 390°C for one hour. The wafer was patterned into 10×10 um$^2$ structure by using optical lithography and etched up to the Ru(20) seed layer by Ion Beam Etching, where the ground electrode is connected to the seed layer. Then the p-MTJs were fully covered by the $SiO_2$ deposited by chemical vapor deposition. The nanocontact size of 100 nm was fabricated using electron beam lithography and inductively coupled plasma etch with optical emission spectrometer. Finally, a Ti(20)/Pt(200) top electrode was fabricated by a lift-off process.

**Devices characterization.** The magnetization of the film was carried out using Vibrating Sample Magnetometer (VSM) instrument. The dc and microwave characterization on the NC-STNO devices was revealed by a custom-built probe station, capable of independently controlling the field magnitude and angle. The dc current was provided by a Keithley 6221 current source. The voltage was measured with a Keithley 2182 nanovoltmeter. The generated microwave signals were decoupled from the dc current via a bias tee, amplified with a 0.1-25 GHz low-noise amplifier and analyzed by an Rohde & Schwarz FSU spectrum analyzer.

**Simulations.** Micromagnetic simulations are performed on the graphics-processing-unit-based tool Mumax3[36]. An NC-STNO geometry is modeled in our simulations. The diameter of an MTJ is 240 nm, while that of the NC above the MTJ is 100 nm (See supplementary note 3). Though the size of the simulated device is much smaller than our sample, we consider it a reasonable simplification to capture the physical picture of our experiments because the current is roughly distributed within the nanocontact area, even when the simulated geometry is enlarged to 1 μm. The free layer is set as 1 nm for the SFL MTJ. In contrast, the DFL MTJ is represented by two ferromagnetically coupled magnetic layers, with a thickness of 0.5 nm for the top layer and 1 nm for the bottom layer. The spacer between the two layers is omitted. A discrete mech of 2 nm × 2 nm × 0.5 nm is used in our simulations.

The magnetization dynamics of each site in the free layer are numerically calculated using the Landau-Lifshitz-Gilbert (LLG) equation with the STT and Zhang-Li torque included (see supplementary note 4). The following magnetic parameters are used: spin polarization ratio $P$ = 0.4, exchange constant $A$ = 15 pJ m$^{-1}$ saturation magnetization $M_s$ = 987 KA m$^{-1}$ from VSM results, Gilbert damping $α$ = 0.0218 and anisotropy energy density $K_u$ = 0.66 × 10$^6$ J m$^{-3}$ from FMR results. If no pinning effect is considered, the DMI strength D is set as 0 and the exchange stiffness between the coupled two free layers is set as *2.5 × 10$^{-14}$ J m$^{-1}$,* which corresponds to an interlayer coupling strength of 0.1 *mJ m$^{-2}$*. When introducing the pinning effect, we set DMI strength D = ± 0.5 *mJ m$^{-2}$* for the two coupled layers, respectively. In the meantime, spatial distribution of interlayer coupling strength from 0.04 to 0.1 *mJ m$^{-2}$* is set using a Voronoi tessellation with an average size of 5 nm. The current distribution in the free layer is simulated by the COMSOL software (see supplementary note 5) and employed as an input in the micromagnetic simulations.

## ASSOCIATED CONTENT

**Supporting Information**

The FMR results, dynamical properties under high magnetic fields, current-field phase diagrams of normalized magnetoresistance, models and details for other micromagnetic simulations, comparison of droplets functions in MTJs, SVs, and SHNOs are reported in the Supporting Information.

## AUTHOR INFORMATION

**Corresponding Author**

*Correspondence author: weisheng.zhao@buaa.edu.cn; johan.akerman@physics.gu.se

**Author contributions**

[†]K. Shi, W. Cai, and S. Jiang contributed equally to this work. W. Z. and J. Å. initialized, conceived and supervised the project. K. S., K. C., Z. G., Y. H., J. W, and J. Y. fabricated the devices. K. S., W. C., S. J., and A. D. performed the measurements. Z. L. grew the films. K. S. and D. Z. performed the micromagnetic simulations. K. S., W. C., S. J., J. Å. and W. Z. analyzed the results and wrote the manuscript. All authors discussed the results and implications.

**Notes**

The authors declare no competing financial interest.

## ACKNOWLEDGMENTS

The authors gratefully acknowledge the National Natural Science Foundation of China

(Grant Nos 61627813, 61904009), the National Key Technology Program of China 2017ZX01032101 and the China Postdoctoral Science Foundation funded project (2018M641151) for their financial support of this work.

**Figures**

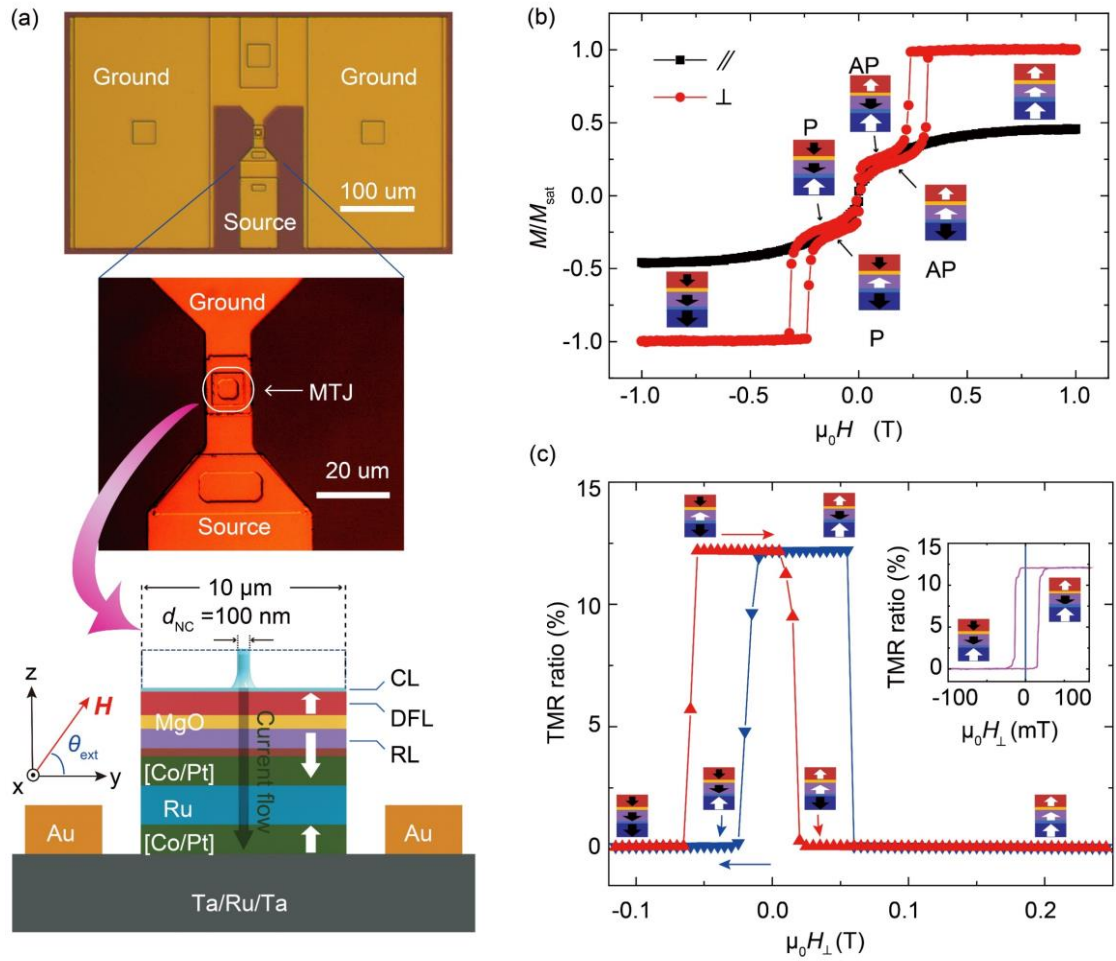

**Figure 1 Device schematic and static characterization. (a)** A schematic of the pMTJ STNO device. The external magnetic field $H$ is applied at an angle $\theta_{ext}$. **(b)**, Magnetization hysteresis loops of the film, normalized to the saturation magnetization $M_{sat}$. **(c)**, TMR loops measured at $I$ = -0.1 mA. Positive current flows from the Source to the Ground.

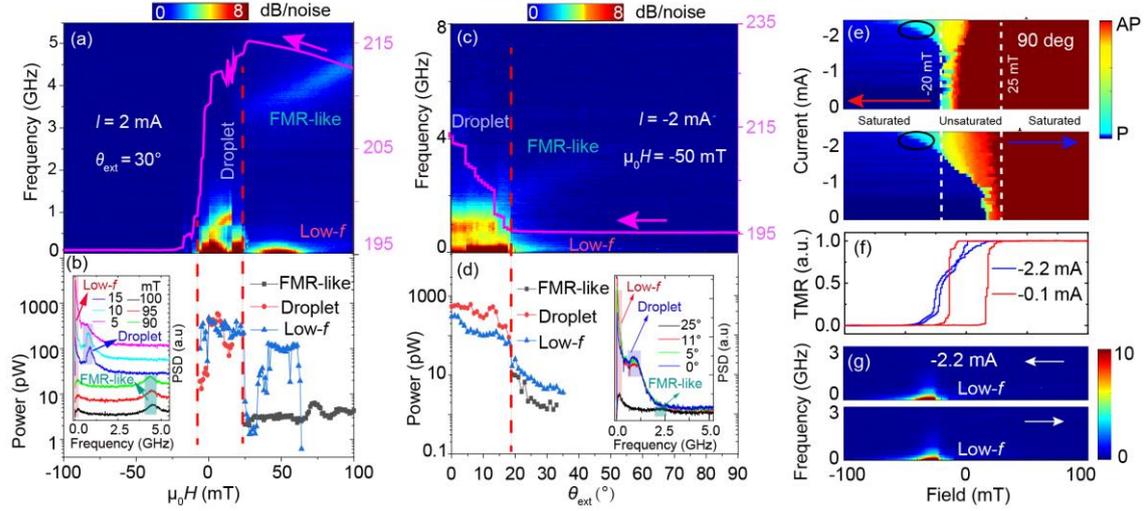

**Figure 2 Dynamical properties of magnetic droplets.** Power Spectral Density (PSD) (left) and resistance (right) as functions of **(a)** magnetic field applied at an angle of 30° and $I$ = 2 mA, and **(c)** angle at $\mu_0 H$ = 20 mT and $I$ = -1.6 mA. The pink solid lines give the resistance curves. The corresponding integrated powers as functions of **(b)** magnetic field and **(d)** angle. The red vertical dotted lines display the critical points in the dynamic and static properties. (e) Perpendicular magnetic field-swept resistance measurements at different negative current $I$. The dashed white lines marked the unsaturated fields for DFL. (f) The corresponding magnetoresistance plots under -0.1 and -2.2 mA. (g) PSD as a function of the perpendicular magnetic field with $I$ = -2.2 mA. The low-$f$ noises are accompanied by the intermediated resistance states marked by the circles in (e).

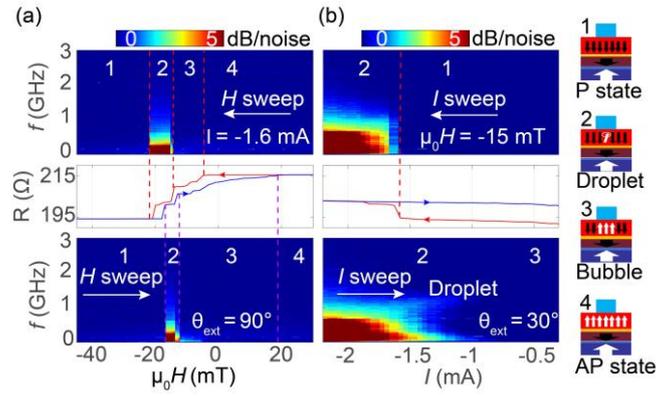

**Figure 3 Nucleation of droplet solitons with field and current sweeps.** Frequency spectra and resistance as functions of **(a)** external perpendicular magnetic field ($I$ = -1.6 mA) and **(b)** current ($\mu_0 H$ = -15 mT, $\theta_{ext}$ = 30°). Right models represent the possible magnetic states: P state, Droplet, Bubble, AP state, named as 1, 2, 3, 4, respectively, corresponding to the MR and PSD results of the NC-STNO.

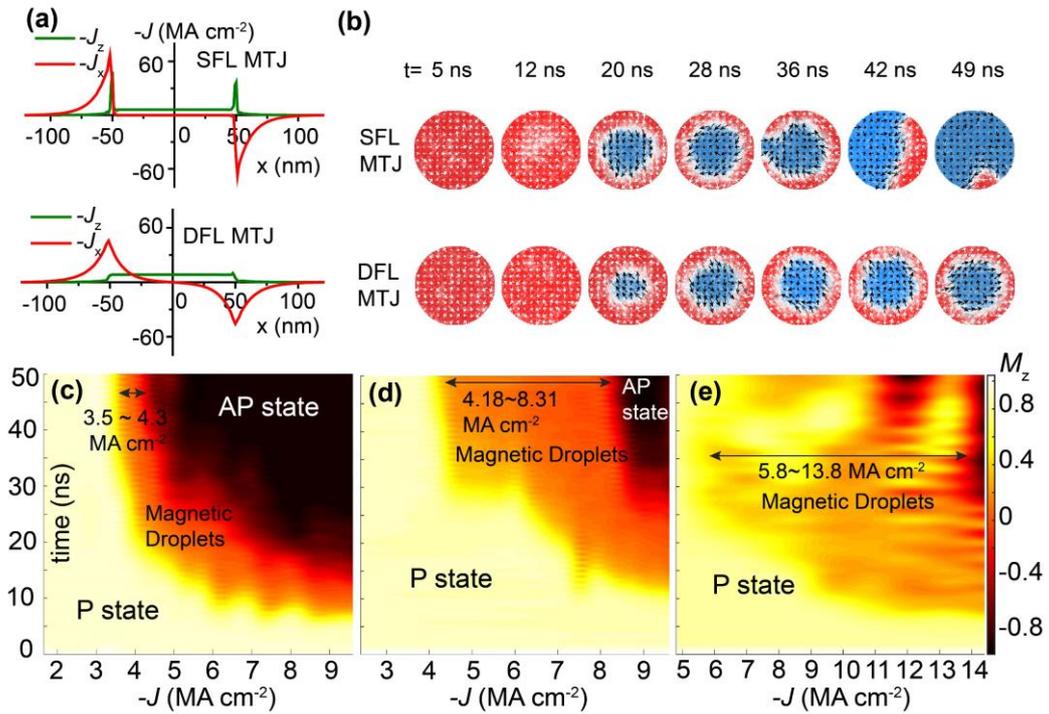

**Figure 4 Micromagnetic Simulation. (a)** The distributions of the current in free layers of the SFL and DFL MTJs. (b) The comparison of the magnetization evolutions under selected time between SFL and DFL where the current density is -7 MA cm$^{-2}$. **(c)-(d)** Plots of current magnitude vs time of the nucleation magnetic droplets in SFL-MTJ, DFL-MTJ, and DFL-MTJ with RKKY and DMI distribution, respectively. The horizontal arrows present the current range for stable magnetic droplets. All the simulations are carried out at -10 mT and 300 K. The initial state for the device is P state.